\documentclass[conference]{IEEEtran}

\usepackage{cite}
\usepackage{amsmath,amssymb,amsfonts}
\usepackage{algorithmic}
\usepackage{graphicx}
\usepackage{textcomp}
\usepackage{xcolor}
\usepackage{booktabs} 
\usepackage{tabularx}
\usepackage{float}

\newcolumntype{L}{>{\raggedright\arraybackslash}X} 
\newcolumntype{C}{>{\centering\arraybackslash}X}   
\newcolumntype{R}{>{\raggedleft\arraybackslash}X}  

\def\BibTeX{{\rm B\kern-.05em{\sc i\kern-.025em b}\kern-.08em
    T\kern-.1667em\lower.7ex\hbox{E}\kern-.125emX}}

\newcommand\numberthis{\addtocounter{equation}{1}\tag{\theequation}}
    
\begin{document}

\title{Group Relative Policy Optimization for Speech Recognition}



\author{\IEEEauthorblockN{Prashanth Gurunath Shivakumar, Yile Gu, Ankur Gandhe, Ivan Bulyko}
\IEEEauthorblockA{\textit{Amazon Science, Seattle, U.S.A} \\
\texttt{prashanth.g.s@ieee.org, \{yilegu,aggandhe,ibbulyko\}@amazon.com}
}
}



\maketitle

\begin{abstract}
Speech Recognition has seen a dramatic shift towards adopting Large Language Models (LLMs). This shift is partly driven by good scalability properties demonstrated by LLMs, ability to leverage large amounts of labelled, unlabelled speech and text data, streaming capabilities with auto-regressive framework and multi-tasking with instruction following characteristics of LLMs. However, simple next-token prediction objective, typically employed with LLMs, have certain limitations in performance and challenges with hallucinations. In this paper, we propose application of Group Relative Policy Optimization (GRPO) to enable reinforcement learning from human feedback for automatic speech recognition (ASR). We design simple rule based reward functions to guide the policy updates. We demonstrate significant improvements in word error rate (upto 18.4\% relative), reduction in hallucinations, increased robustness on out-of-domain datasets and effectiveness in domain adaptation.
\end{abstract}


\section{Introduction}
Recently, significant strides have been made in Automatic Speech Recognition by adoption of large language models based on causal, decoder-only, transformer architectures.
This is mainly driven by remarkable scaling properties of the LLMs with the ability to leverage large amounts of supervised, unsupervised speech and text data, streaming friendly properties due to causality imposed on transformers, and its simplicity in terms of less components which enables to treat the models as a black-box.

Typical auto-regressive LLM models the probability of the next token given a sequence of tokens.
This paradigm can be extended to include speech modality by modeling the probability of next text token conditioned on a sequence of audio representations or audio tokens.
\cite{lakhotia2021generative} proposed to use acoustic tokens derived from K-means quantized HuBERT embeddings to model speech continuation with LLMs.
VoxtLM \cite{maiti2023voxtlm}, SpeechGPT \cite{zhang2023speechgpt} used such discrete audio units in application to ASR in a multi-task framework.
Several studies \cite{du2023lauragpt, ma2024embarrassingly, wu2023decoder, chu2023qwen, tang2023salmonn, yu2024connecting, fathullah2024prompting} demonstrated strong ASR performance with directly feeding continuous speech embeddings to LLM.
Conformer derived embeddings \cite{du2023lauragpt, wu2023decoder, fathullah2024prompting}, HuBERT embeddings \cite{li2023prompting}, Whisper encoder \cite{ma2024embarrassingly, chu2023qwen, tang2023salmonn, yu2024connecting}, WavLM \cite{ma2024embarrassingly} are popular speech representations employed.
While some studies have explored fixed conformer embeddings \cite{wu2023decoder} with learnable speech projection layers, others choose to update conformer parameters during LLM training \cite{li2023prompting, chu2023qwen}.
\cite{wu2023decoder, tang2023salmonn, fathullah2024prompting} utilized low rank adapters (LoRA) for fine-tuning the LLM.
Some prior studies \cite{chu2023qwen} have adopted 2 stage training, which comprises freezing LLM while updating the audio encoder in the first stage. Second stage involves fixing the audio encoder and fine-tuning the LLM.
\cite{li2023prompting} proposed a deep fusion mechanism based on gated cross attention modules operating on HuBERT features. The HuBERT model parameters were also updated during the LLM training.
\cite{yu2024connecting} conducted detailed study on three speech to LLM interface modules, including simple linear projection, multi-head cross-attention and Q-Former modules and found the latter to be better for ASR.

While the LLM based ASR has made significant strides in improving recognition rate, they often suffer with hallucinations \cite{atwany2025lost, frieske2024hallucinations}.
Hallucinations are characterized with high insertion rates and arise when the model deviate from input stimuli (speech signal), prioritizing distributional patterns resulting in semantically and phonetically unrelated outputs.
Such hallucinations can have dangerous impacts, including deception, on applications in domain where high precision is critical.
\cite{atwany2025lost} conducted a detailed study characterizing the LLM hallucinations in the context of ASR.
They find that low WER can often conceal big hallucinations, and the hallucinations can result from common audio signal perturbations including noise, pitch shift, temporal shifts.
Further \cite{frieske2024hallucinations} finds that noise in human annotations, labeling, can lead to hallucinations.

One plausible way to further optimize speech recognition and increase its robustness to hallucinations is with re-inforcement learning (RL).
Several studies have explored re-inforcement learning techniques in application to speech recognition.
Early applications of re-inforcement learning attempted to correct errors in isolated word recognition \cite{lee1990corrective}.
A confidence based RL scheme based on incremental conditional entropy maximization was proposed in \cite{molina2009maximum} demonstrating reduction in WER up-to 18\%.
Policy gradient based approach was explored in \cite{tjandra2019end} which allows to directly optimize the edit-distance resulting in up-to 6\% improvement in character error rate (CER).
\cite{kala2018reinforcement} applied re-inforcement learning in a hypothesis selection framework over the n-best model output.
Authors in \cite{chen2023leveraging}, proposed re-inforcement framework for fusing multiple modalities, i.e., speech and video for improving auto-regressive audio visual speech recognition.

In the context of LLMs, recent strides have been made in re-inforcement learning from human feedback (RLHF).
OpenAI in \cite{schulman2017proximal} introduced a family of RL algorithms based on policy gradient methods named proximal policy optimization (PPO).
PPO paved a path towards a stabile, feasible and efficient way to enable RLHF for LLMs.
Recently, \cite{shao2024deepseekmath} proposed Group Relative Policy Optimization (GRPO) for preference optimization of LLMs.
GRPO simplifies PPO by dropping the critic model and instead employs average of rewards for advantage estimation.
In its first application \cite{shao2024deepseekmath}, the GRPO provided accuracy gains on math benchmarks.
Subsequently, the algorithm provided a framework for building reasoning capabilities for LLMs \cite{guo2025deepseek}.
Dynamic sampling policy optimization (DAPO) further tweaked the GRPO to enhance training effectiveness by introducing practical tricks including decoupled clip, token-level policy gradient loss and dynamic sampling \cite{yu2025dapo}.
\cite{liu2025understanding} proposed a variant of GRPO named Dr. GRPO improving the token efficiency by making the optimization unbiased.

However, there have been few attempts at applying RLHF for LLMs operating on speech.
\cite{lin2024align} first proposed application of DPO to enhance speech continuation on spoken language models by introducing AI experts for generating automated preference data.
Qwen2-Audio \cite{chu2024qwen2} applied DPO with human preferences to optimize speech understanding.
Qwen2.5-Omni \cite{xu2025qwen2} applied DPO to enhance speech generation.
Omni-R1 \cite{zhong2025omni} further extended Qwen2.5-Omni model's capabilities by applying GRPO.
\cite{zhao2025r1} applied GRPO to improve on emotion recognition and reasoning capabilities that enables the model to better analyze visual and audio modalities.
\cite{rouditchenko2025omni, wen2025sari, li2025reinforcement} fine-tuned Qwen2.5-Omni on audio QA datasets using GRPO to achieve state-of-the-art performance on MMAU and MMAR benchmarks.
\cite{sun2025f5r} applied GRPO for improving text-to-speech system by designing rewards to optimize the TTS-WER and speaker similarity of the synthesized speech.

In this study, we propose application of GRPO towards LLM based ASR as an additional stage of fine-tuning towards optimizing the overall performance and improve robustness of the system to hallucinations in application to out-of-domain, unseen acoustic conditions.
To the best of our knowledge, this is the first attempt at application of RLHF to LLMs to improve speech recognition.
We design and explore various reward functions and present our findings and strategies to improve speech recognition with re-inforcement learning.
The rest of the paper is organized as follows: Section~\ref{sec:technique} presents the proposed LLM based ASR system.
Section~\ref{sec:setup} provides the description of our experimental setup and datasets employed in our study.
Section~\ref{sec:results} presents the experimental results and the discussions.
Finally, the study and its findings are concluded in Section~\ref{sec:conclusion}.

\section{Proposed Technique}\label{sec:technique}
\subsection{Auto-regressive LLM based Speech Recognition}
Auto-regressive, causal, decoder only LLMs model the probability of the next token given a sequence of tokens:
\begin{equation}\label{eq:1}
P_{LM}(X) = \prod_{t=1}^{T} P(x_t|x_{t-1},\ldots,x_1)
\end{equation}
where $x_t \in V_{txt}$ is a text token belonging to text vocabulary $V_{txt}$.
Such a model can be adopted to the task of speech recognition using next token paradigm by modeling:
\begin{equation}\label{eq:2}
P_{ASR}(X|S) = \prod_{t=1}^{T} P(x_t|x_{t-1},\ldots,x_1, s_N,\ldots,s_1) 
\end{equation}
where $s_1,\ldots,s_N$ are acoustic units or representations of length $N$ frames corresponding to their transcriptions $x_1,\ldots,x_T$ of length $T$.
In this work, $s_t$ are continuous vector representations derived from a pre-trained acoustic encoder.
The details of acoustic encoder are described under section~\ref{sec:setup}.
A simple linear projection, feed-forward layer is used as interface in mapping the acoustic representation to LLM input.
The acoustic encoder is frozen during the training, while both the linear projection and LLM parameters are updated.
In case of ASR, we compute the next-token prediction loss only on the output transcriptions.
The model is trained in 2 stages.
During the first stage, LLM is pre-trained on large corpus of text using Equation~\ref{eq:1}.
Next, the LLM is fine-tuned on parallel speech-text supervised data using Equation~\ref{eq:2}.
The prompt format for ASR in our setup comprises \texttt{<User><BOS> Convert speech to text <S-BOS> $s_1, s_2, \ldots, s_N$ <System> <BOS> $x_1, x_2, \ldots, x_T$ <EOS>}.
During inference, the LLM is prompted with \texttt{<User><BOS> Convert speech to text <S-BOS> $s_1, s_2, \ldots, s_N$ <System>}.
In the above sequences, \texttt{<User>, <BOS>, <S-BOS>, <EOS>} are special tokens.

\subsection{Group Relative Policy Optimization for Speech Recognition}
In this study, we propose an additional fine-tuning stage based on RLHF for further performance optimization and robustness.
RLHF algorithms provide an effective means of incorporating human preferences to direct LLM generations.
PPO introduced in \cite{schulman2017proximal} is a popular policy gradient RL technique for LLMs.
PPO is based on the Actor-Critic model, where the Actor interacts with the environment collecting rewards associated with each action which pertain to quantifying how good or bad was the action taken.
A critic model is trained alongside to estimate the expected future reward from the current state.
The PPO then computes advantages for each action taken which describes the quality of the action relative to the average expected return.
More importantly, PPO paved a path towards a stable, feasible and efficient way to enable RLHF for LLMs.
Group Relative Policy Optimization (GRPO) introduced in \cite{shao2024deepseekmath}, is a variant of PPO, primarily designed for preference optimization of LLMs.
GRPO simplifies PPO by dropping the critic model and instead employs average of rewards for advantage estimation.

In the context of ASR, such RL techniques can directly optimize the speech recognition output to human transcriptions.
While, technically, both PPO and GRPO can be applicable, GRPO provides simpler framework to achieve our objectives.
Further, in the case of ASR, the human feedback is derived directly from groundtruth transcripts, hence, objective, rule-based rewards can suffice.
Thus, GRPO is a better fit without necessitating the need for training separate reward models.
The reward models themselves often suffer from reward hacking problem \cite{everitt2017reinforcement}.
Moreover studies such as \cite{shao2024deepseekmath}, have found GRPO to yield similar performance gains as PPO.

The GRPO optimizes and maximizes the following objective:
\resizebox{\linewidth}{!}{
\begin{minipage}{\linewidth}
\begin{align*}
L_{\scriptscriptstyle{GRPO}}=\frac{1}{G} \sum_{i=1}^G \frac{1}{|o_i|}\sum_{t=1}^{|o_i|} min[\pi_{\vartheta} \hat{A}_{i,t}, & clip(\pi_{\vartheta}, 1-\varepsilon, 1+\varepsilon]) \hat{A}_{i,t}] \\& - \beta \mathbb{D}_{KL}[\pi_{\theta}||\pi_{ref}] \numberthis \label{equation:loss}
\end{align*}
\end{minipage}
}

where $G$ is the number of generations, $o_i$ is the $i^{th}$ generated output hypothesis, $\hat{A}_{i,t}$ is the advantage, given by:
\begin{equation}
\hat{A}_{i,t} = \frac{\mathbb{R}(i) - \mathbb{E}[\mathbb{R}(i)]}{\sigma(R(i))}
\end{equation}
where $\mathbb{R}(i)$ is the reward for generated output $o_i$, $\mathbb{E}[R(i)]$ is the expected reward across $G$ generations and $\sigma(R(i))$ is the standard deviation.
$\varepsilon$ is the parameter introduced in PPO \cite{schulman2017proximal} for clipping and stabilizing the training, $\beta$ is a hyper-parameter that controls the deviation of the policy from the reference seed model, $\mathbb{D}_{KL}[\pi_{\theta}||\pi_{ref}]$ is the Kullback-Leibler divergence between the reference model $\pi_{ref}$ and current policy model $\pi_{\theta}$ given by:
\begin{equation}
\mathbb{D}_{KL}[\pi_{\theta}||\pi_{ref}] = \frac{\pi_{ref}(o_{i,t}|o_{i,<t}, s)}{\pi_{\theta}(o_{i,t}|o_{i, <t}, s)} - log \frac{\pi_{ref}(o_{i,t}|o_{i,<t}, s)}{\pi_{\theta}(o_{i,t}|o_{i, <t}, s)} - 1
\end{equation}
and $\pi_{\vartheta}$ is given by:
\begin{equation}
\pi_{\vartheta} = \frac{\pi_{\theta}(o_{i,t}|o_{i,<t}, s)}{\pi_{\theta_{old}}(o_{i,t}|o_{i, <t}, s)}
\end{equation}
where $\pi_{\theta_{old}}$ is the old policy model as defined in PPO \cite{schulman2017proximal}.

Additionally, we also take learnings from DAPO \cite{yu2025dapo}, Dr. GRPO \cite{liu2025understanding} and explore its effect in this study.
DAPO enforces three key modifications to the loss, (i) introduces an upper clipping threshold to increase probability of unlikely exploration, (ii) remove KL-divergence term ($\beta = 0$), and (iii) token-level policy gradient loss computation (versus sample-level) to account for biases introduced as a function of sample length.
On the other hand, Dr. GRPO proposes two key modifications to loss towards unbiased optimization, (i) removal of standard deviation normalization in advantage computation, and (ii) length normalization of the GRPO loss.
These modifications are geared towards preventing model's bias towards longer, incorrect responses.

\subsection{Rule-based Rewards for Speech Recognition}
Given that the human feedback for speech recognition is via ground-truth human transcriptions, we propose to use simple rule-based rewards.
This simplifies the setup, reduces computational complexity and helps avoid reward hacking problem \cite{everitt2017reinforcement}.
In this study, we explore the following rewards:\\
\textbf{Word Error Rate (WER)}: Negated word-error-rate can serve as a potential reward and helps optimize directly to the target metric. WER is a normalized version of the edit-distance, hence can reinforce the model without any biases towards the length of the audio.
\begin{equation}
\mathbb{R}_i = -WER = -\frac{Sub + Del + Ins}{N}
\end{equation}
where $Sub$, $Del$, $Ins$ are substitutions, deletions and insertions respectively, derived from dynamic alignment, $N$ is the total words in the reference.\\
\textbf{Exact Match (EM)}: Several studies \cite{yu2025dapo} recommend simpler rewards such as exact match which is proven to be an effective approach for invoking reasoning capabilities.
\begin{equation}
\mathbb{R}(ref, hyp) = 
\begin{cases}
1 & \text{if  } ref = hyp \\
0 & otherwise
\end{cases}
\end{equation}
In context of ASR, exact match is invariant of the sample length as well as the number of errors.\\
\textbf{Total Errors (ED)}: Number of incorrect recognition can be used as a reward. This is equivalent to un-normalized edit distance over word sequence between reference and generated candidates.
\begin{equation}
\mathbb{R}_i = -(Sub + Del + Ins)
\end{equation}
This enables the optimization to concentrate towards samples that are drastically different to reference.

\section{Data and Experimental Setup}\label{sec:setup}
\begin{table}[!b]
    \centering
    \caption{Speech Datasets Statistics}
    \label{table:data}
    \begin{tabularx}{0.8\columnwidth}{@{} p{\dimexpr.6\linewidth-2\tabcolsep-1.3333\arrayrulewidth} L @{}} 
        \toprule
        \textbf{Datasets} & \textbf{Hours} \\
        \midrule
        Fleurs \cite{conneau2023fleurs} & 987 hrs \\
        Multi-lingual Librispeech \cite{pratap2020mls} & 50k hrs \\
        Voxpopuli \cite{wang2021voxpopuli} & 1791 hrs \\
        People Speech \cite{galvez2021people} & 30k hrs  \\
        Common Voice \cite{ardila2019common} & 2k hrs \\
        Covost2 \cite{wang2020covost} & 3.5k hrs \\
        \bottomrule
    \end{tabularx}
\end{table}

\subsection{Experimental Setup}
In this work we employ two LLM models based on Llama3 architecture \cite{grattafiori2024llama}, (i) smaller, 2B parameter model, and (ii) larger, 8B parameter model.
The 2B model comprises 24 layers, 16 attention heads, hidden dimension of 2048 and feed-forward dimension of 8192.
The 8B model comprises of 32 layers, 32 attention heads, hidden dimension of 4096 and feed-forward dimension of 16384.
The 8B model employs weight tying between the embedding and output layers.
Both the LLMs use 8 query groups, SwiGLU activation and has a vocabulary size of 187178.

For consuming speech, our setup employs a speech encoder module based on \cite{male2025durepdualmodespeechrepresentation}.
The encoder is primarily based on the conformer architecture with 2B parameters.
The encoder has a frame rate of 40ms and a hidden dimension of 2048, more details are available in \cite{male2025durepdualmodespeechrepresentation}.
A weighted combinations of multiple layers are used to encode speech signals onto 4096 dimensional embeddings.
A linear projection layer is used to map the 4096 dimension embedding to the embedding dimension of the LLMs.

\begin{table*}[!t]
    \centering
    \caption{Experimental Results: Word-Error-Rate Metric (Acronyms in parenthesis, WER, EM, ED, correspond to reward functions)\\
    SFT: Supervised Fine-Tuning; GRPO: Group Relative Policy Optimization; DAPO: Dynamic Sampling Policy Optimization}
    \label{table:results}
    \begin{tabularx}{\textwidth}{@{} p{\dimexpr.15\linewidth-2\tabcolsep-1.3333\arrayrulewidth} C C C C C C C C C C C C C @{}} 
        \toprule
        \textbf{Datasets} & \textbf{People-Speech} & \multicolumn{6}{c}{\textbf{Multilingual Librispeech}} & \multicolumn{6}{c}{\textbf{Voxpopuli}} \\
        \textbf{Language} & En & En & Fr & It & De & Es & Overall & En & Fr & It & De & Es & Overall \\ 
        \midrule
        2B SFT & 23.5 & 4.83 & 5.35 & \textbf{10.5} & 6.06 & 3.46 & 5.48 & 7.92 & 9.39 & 15.83 & 10.95 & 7.64 & 10.12 \\
          + GRPO (WER) & 21.53 & 4.85 & 5.22 & 10.7 & 6.15 & 3.43 & 5.49 & 7.82 & 8.5 & 14.93 & 9.89 & 7.24 & 9.47 \\
          \hspace{0.5cm}+ $\beta=0$ & \textbf{21.3} & 4.92 & 5.37 & 11.32 & 6.15 & 3.4 & 5.59 & 7.96 & 8.56 & 14.53 & \textbf{8.96} & \textbf{7.14} & 9.24 \\
          + GRPO (ED) & \textbf{21.48} & 5.19 & 5.39 & 10.7 & \textbf{6.05} & 3.66 & 5.64 & 7.87 & 8.55 & 14.91 & 9.86 & 7.38 & 9.51 \\
          + GRPO (EM) & 22.29 & 4.93 & 5.48 & 11.06 & 6.2 & 3.47 & 5.62 & 7.88 & 8.61 & 15.98 & 10.18 & 7.32 & 9.76 \\
          + DAPO (EM) & 21.71 & \textbf{4.79} & 5.24 & 11.54 & 6.13 & \textbf{3.24} & 5.51 & 7.98 & 8.53 & 15.14 & 9.81 & 7.43 & 9.57 \\
          + DAPO (WER) & \textbf{21.30} & 5.04 & \textbf{5.2} & 10.64 & 6.21 & 3.32 & 5.53 & \textbf{7.8} & 8.57 & \textbf{14.42} & 9.39 & 7.27 & 9.30 \\
          + DR-GRPO (EM) & 22.24 & 4.91 & 5.27 & 11.02 & 6.08 & 3.63 & 5.56 & 7.87 & \textbf{8.48} & 15.16 & 10.58 & 7.49 & 9.70 \\
        \midrule
        8B SFT & 25.48 & 4.46 & 4.87 & 9.37 & \textbf{5.07} & 3.08 & 4.87 & \textbf{7.5} & 9.21 & 14.66 & 10.9 & 7.05 & 9.66 \\
          + GRPO (WER) & \textbf{21.42} & 4.64 & \textbf{4.7} & 9.45 & 5.33 & \textbf{3.02} & 4.95 & 7.66 & 8.6 & \textbf{14.04} & \textbf{8.89} & \textbf{6.64} & 8.98 \\
          + GRPO (EM) & 22.49 & \textbf{4.41} & 4.86 & \textbf{9.16} & 5.25 & 3.03 & 4.87 & 7.66 & \textbf{8.42} & 14.93 & 9.34  & 7.03 & 9.27 \\
        \bottomrule
    \end{tabularx}
\end{table*}

For speech recognition, the LLM is first pre-trained on text-only data with a constant learning rate schedule of 1e-4 using Adam optimizer.
The global batch size is approximately 1M tokens with max-sequence length set to 2048.
Motivation to pre-train on text is derived from \cite{hassid2024textually, nguyen2025spirit, maiti2024voxtlm}, which have shown benefits for speech related tasks.
Next, the LLM is fine-tuned using speech data with a cosine learning rate scheduler with peak learning rate of 5e-6 for 100k steps, with 1000 step warm-up.
The global batch size is set to 128 and sequence length is capped at 2048 tokens.

GRPO training stage uses the fine-tuned model as the reference.
It comprises generating $G$ \texttt{<System>} responses when prompted with the \texttt{<User>} sequence to compute the loss in Equation~\eqref{equation:loss}.
GRPO is conducted with a fixed learning rate of 1e-6 similar, global batch size of 64 for a maximum of 5000 steps.
We experiment with different generation configurations including $\beta$, number of generations, generation strategies, reward scaling.
For all of our experiments, the top\_k and min\_p is set to None, top\_p = 1.0, repetition penalty = 1.0, $\beta$ = 0.04 (when used).

\subsection{Data}
The pre-training text data is based on RedPajama \cite{together2023redpajama}.
The speech datasets used in this study is presented in Table~\ref{table:data}.
We use open-sourced multi-lingual speech datasets comprising approximately 88,000 hours of training data for supervised fine-tuning as well as GRPO.
The datasets are mixed with weights representative of their sizes, emphasizing English, French, Spanish, Italian and German.
We maintain held-out development and evaluation partitions for each of the corpus.
Best model checkpoints are picked based on development partition.
We present evaluations on People's speech, multilingual librispeech and Voxpopuli datasets to capture diverse range of variability.
We also provide a language level breakdown of WER for MLS and Voxpopuli evaluations.
Additionally, we use AMI corpus \cite{carletta2005ami}, and TEDLium \cite{hernandez2018ted} for out-of-domain evaluations.

\section{Results}\label{sec:results}

Table ~\ref{table:results} presents the experimental results for 2B and 8B models on people's speech, multi-lingual librispeech and voxpopuli.
We present detailed results with different configurations on the 2B model.
The 8B model is used to assess the effect of scaling on the proposed method.
Firstly, comparing the 2B models, it is evident that the proposed method improves over the reference SFT model on most languages across datasets.
We observe up-to 8.6\% relative on people's speech, up-to 5.8\% relative on multilingual librispeech  (MLS) and up-to 18.2\% relative improvements on Voxpopuli.
Assessing the results with the 8B model, firstly we note the improvements with the bigger model in comparison to 2B SFT model (with the exception of people's speech).
Comparisons of the 8B SFT model with the GRPO, paint a similar picture as 2B, with up-to 15.7\% relative improvements on people's speech, up-to 3.5\% relative improvement on MLS and 18.4\% relative improvement on Voxpopuli.
Degradations, if any, are small, demonstrating the robustness of the proposed technique across varying acoustic conditions.

\begin{table*}[!t]
    \centering
    \caption{Experimental Results: Out-of-domain Evaluations. WER and its breakdown in Insertions, Deletions and Substitutions}
    \label{table:ood}
    \begin{tabularx}{\textwidth}{@{} L C C C C C C @{}}
        \toprule
        \textbf{Datasets} & \multicolumn{2}{c}{\textbf{TEDLIUM}} & \multicolumn{2}{c}{\textbf{AMI-IHM}} & \multicolumn{2}{c}{\textbf{AMI-SDM}} \\
        \textbf{Models} & Ins / Del / Sub & WER & Ins / Del / Sub & WER & Ins / Del / Sub & WER \\
        \midrule
        2B SFT & 0.5 / 1.7 / 1.6 & 3.9 & 38.4 / 10.3 / 7.3 & 56.0 & 56.7 / 16.6 / 14.6 & 87.88 \\
        + GRPO (WER) & 0.4 / 1.6 / 1.6 & 3.7 & 2.8 / 10.96 / 6.3 & 20.0 & 7.4 / 18.1 / 12.1 & 37.59 \\
        + GRPO (ED) & 0.5 / 1.7 / 1.6 & 3.74 & 5.6 / 11 / 6.3 & 22.84 & 8.9 / 17.6 / 12.1 & 38.52 \\
        + GRPO (EM) & 0.4 / 1.8 / 1.7 & 3.93 & 2.7 / 11.1 / 6.3 & 19.95 & 7.4 / 18.1 / 12.1 & 37.59 \\
        \midrule
        8B SFT & 0.5 / 1.2 / 1.8 & 3.53 & 82 / 10.5 / 7.7 & 100.26 & 195.4 / 17.4 / 15.3 & 227.98 \\
        + GRPO (WER) & 0.4 / 1.28 / 1.67 & 3.36 & 4.6 / 11.6 / 6.0 & 22.16 & 9.0 / 20.2 / 10.5 & 39.69 \\
        \bottomrule
    \end{tabularx}
\end{table*}

\begin{table*}[t]
    \centering
    \caption{Experimental Results in WER: Domain Adaptation.}
    \label{table:domain_adapt}
    \begin{tabularx}{0.89\textwidth}{lcccccccccccc}
        \toprule
        \textbf{Datasets} & \textbf{AMI-IHM} & \textbf{AMI-SDM} & \multicolumn{5}{c}{\textbf{Multilingual Librispeech}} & \multicolumn{5}{c}{\textbf{Voxpopuli}}\\
        \textbf{Language} & En & En & En & Fr & It & De & Es & En & Fr & It & De & Es \\ 
        \midrule
        2B Baseline & 56.0 & 87.88 & \textbf{4.83} & \textbf{5.35} & \textbf{10.5} & \textbf{6.06} & \textbf{3.46} & \textbf{7.92} & \textbf{9.39} & \textbf{15.83} & \textbf{10.95} & \textbf{7.64} \\
        \midrule
        AMI-SFT & 19.17 & 44.34 & 6.91 & 12.10 & 14.88 & 8.39 & 7.57 & 9.80 & 12.48 & 19.33 & 12.92 & 12.21 \\
        AMI-GRPO (WER) & \textbf{15.55} & \textbf{31.98} & 6.59 & 10.34 & 15.80 & 9.57 & 5.90 & 9.10 & 12.15 & 19.40 & 14.51 & 10.62 \\
        \bottomrule
    \end{tabularx}
\end{table*}

Next, we assess different configurations of the proposed techniques:\\
\textbf{Role of Rewards:} We compare the 3 rewards as described under section~\ref{sec:technique}. We found that the WER and the total error rewards fare better when the absolute WER of the datasets are high. The exact-match performs relatively poor in high WER conditions likely due to lower probability of generating outputs that positively match the reference. However, the exact-match performs on-par with the WER reward on datasets where the absolute WER is low. On the other hand, the total errors (ED) exhibits an opposite trend. Overall, the WER based reward strikes a better balance across varying WER conditions.\\
\textbf{Role of KL Divergence:} $\beta$ regulates the divergence of the policy model from the reference model. A $\beta$ of 0 removes any regulation and essentially removes the KL Divergence term. In our experiments we do not observe significant divergence in WER when $\beta=0$.\\
\textbf{Role of RL Algorithms}: We compare GRPO, DAPO and Dr. GRPO with exact-match reward. The results suggest that both DAPO and Dr. GRPO outperform traditional GRPO in most cases.\\
\textbf{Role of Generation Strategies}: Beam search decoding and multi-nomial sampling decoding strategies were explored. We found that beam search often leads to better improvements on noisy datasets like people speech and the multinomial sampling offers better improvements on cleaner datasets with lower WER.\\
\textbf{Number of Generations:} In our experiments we tested $G=\{6,10\}$. However, we found the impact to be insignificant and hence skip the results.

\subsection{Out-of-Domain Performance Evaluations}
Translation of reliable performance to unseen acoustic environments is critical for any robust ASR system.
Particularly, in case of auto-regressive LLMs, this often leads to hallucinations especially in noisier acoustic environments, characterized with reverberations, overlapping speech and background noises.
To assess the performance of the proposed technique, we conduct evaluations on unseen datasets including TEDLIUM and AMI meeting corpus.
TEDLIUM comprises of TED-talks with diverse speakers that help us probe on speaker related challenges including diverse range of accents, fluency. 
AMI meeting corpus poses challenges with far-field speech, overlapped speech and noise. 
Note, both TEDLIUM and AMI meeting corpus are not incorporated during SFT and subsequent GRPO training.

Table ~\ref{table:ood} presents the results on out-of-domain evaluations.
The 2B SFT model, gives a WER of 3.85\% on TEDLIUM, however, performs poorly on AMI, i.e., 55.96\% on IHM and 87.88\% on SDM.
A deeper inspection of the errors on AMI corpus in terms of insertions, deletions and substitutions reveal that insertions dominate the errors hinting towards hallucinations.
After GRPO, we see a dramatic reduction in insertions which drives significant improvements in WER.
It is also important to note that there is significant reduction in substitutions after GRPO.
In case of the 8B model, we observe the 8B SFT model scales well on TEDLIUM with reduction in baseline WER over 2B SFT model.
However, the WER explodes to greater than 100\% which is suggestive of increased hallucinations. 
It is likely that bigger models adapt well to in-domain data and as a consequence worsens hallucinations on unseen acoustical environments.
After GRPO, we see drastic reduction in insertions and substitutions similar to the 2B models.
The results highlight that GRPO can increase the robustness of the LLM and reduce hallucinations.
More importantly the learning extends to unseen datasets and acoustic conditions.

\subsection{Domain Adaptation}
Speech signals are characterized by high variability in multiple domains including speaker environment, noise, room characteristics, reverberation, recording conditions, speaker variability spanning linguistics, accents, fluency, age, and gender.
It is typical to adapt ASR models to unseen domains to optimize performance.
In case of ASR-LLMs, one straightforward option is to fine-tune on new domain.
We evaluate the proposed GRPO training as an alternative and assess the overall robustness.
We start from the 2B SFT model trained on data presented under Table~\ref{table:data} as the baseline (corresponding to row 1 in Table~\ref{table:results}).
We train 2 candidate model on AMI speech corpus as a new domain: (i) SFT adaptation, (ii) GRPO adaptation to assess effectiveness of SFT versus proposed method for domain adaptation.
The choice of AMI speech corpus is due to its distinct characteristics in terms of acoustic environment (supported by results in Table~\ref{table:ood}).
The results are presented under Table~\ref{table:domain_adapt}.
Along with results on AMI, we also provide the results on Multi-lingual librispeech and Voxpopuli to assess the performance trade-off after adaptation.
From the results, it is clear that the baseline model performs poorly on the out-of-domain AMI data.
After adapting the baseline on AMI data using typical next-token prediction SFT, we observe substantial improvements, 66\% relative WER reduction on AMI-IHM and 49\% WER reduction on AMI-SDM subsets.
Meanwhile, we also observe degradations on MLS and Voxpopuli across all languages.
Looking at the results with the proposed GRPO adaptation, we see significant improvement relative to both the baseline (72\% on AMI-IHM and 63.61\% on AMI-SDM) as well as SFT adapted model, i.e., 18\% relative WER reduction on AMI-IHM and 27.9\% on AMI-SDM.
Notably, we observe degradations on MLS and Voxpopuli, however, the degradations are relatively lower compared to the SFT adapted model.
This suggests that the proposed GRPO is a better tool to use for adapting the model to a new unseen data or domain.

A highlight of the above results on AMI is that the proposed GRPO training even without inclusion of AMI datasets, i.e., out-of-domain results in Table~\ref{table:ood}, row 2, gives better results compared to the SFT model adapted on AMI.
This concretely establishes the robustness benefits obtained using proposed method on out-of-domain datasets.

\section{Conclusion}\label{sec:conclusion}
In this work, we propose an additional RLHF training stage for LLM based ASR models using GRPO.
We propose 3 simple rule-based rewards for GRPO to facilitate performance improvements and robustness.
We carefully design experiments to evaluate the performance benefits, assess the robustness of the model to hallucinations, out-of-domain datasets.
Further, we demonstrate the proposed method as an effective tool for domain adaptation purposes.
The experiments demonstrate significant WER reductions obtained using the proposed method.
We also show that the resultant model performs drastically better on out-of-domain datasets that are otherwise prone to hallucinations.
Additional experiments support the viability of the proposed method as an effective model adaptation tool.
We provide detailed discussions on the role of different hyper-parameter settings and present strategies and recommendation for effective usage.

In future, interesting rewards can be designed for specific applications, for example, improve slot-error-rate in spoken language understanding applications, or semantic measures to facilitate better, semantically aligned speech recognition outputs.
This also opens up possibilities in responsible AI domain in censoring certain ASR outputs.
Overall, the proposed method opens up possibilities in aligning and controlling certain aspects of ASR system.

\bibliographystyle{IEEEtran}
\bibliography{refs}

\end{document}